\pdfoutput=1
\documentclass[oldversion,pdftex]{aa}
\usepackage{txfonts}
\usepackage{graphicx}
\usepackage{natbib}
\bibpunct{(}{)}{;}{a}{}{,} 
\begin{document}
\title{Redshifted emission lines and radiative recombination continuum
from the Wolf-Rayet binary $\theta$ Muscae: evidence for a triplet system?}

\subtitle{}

\author{Yasuharu Sugawara\inst{1}
, Yohko Tsuboi\inst{1}
\and
Yoshitomo Maeda\inst{2}
}
\institute{
Department of Physics, Faculty of Science and Engineering, Chuo
University, 1-13-27 Kasuga, Bunkyo-ku, Tokyo 112-8551, Japan
\and
Department of High Energy Astrophysics, Institute of Space and
Astronautical Science (ISAS), Japan Aerospace Exploration Agency (JAXA)
3-1-1 Yoshinodai, Sagamihara, Kanagawa 229-8510, Japan }
\date{Received 21 December 2007 / Accepted 30 July 2008}

\abstract
{ We present \textit{XMM-Newton} observations of the WC binary $\theta$
Muscae (WR 48), the second brightest Wolf-Rayet binary in optical
wavelengths. The system consists of a short-period (19.1375 days)
WC5/WC6 + O6/O7V binary and possibly has an additional O supergiant
companion (O9.5/B0Iab) which is optically identified at a separation of $\sim46$
mas. Strong emission lines from highly ionized ions of C, O, Ne, Mg,
Si, S, Ar, Ca and Fe are detected. The spectra are fitted by a
multi-temperature thin-thermal plasma model with an interstellar
absorption $N_{\mathrm{H}} =$ 2--3 $\times 10^{21}$
$\mathrm{cm}^{-2}$. Lack of nitrogen line indicates that the abundance of carbon is at least an order of
magnitude larger than that of nitrogen. A Doppler shift of
$\sim630$ $\mathrm{km}$ $\mathrm{s}^{-1}$ is detected for the O
V\hspace{-.1em}I\hspace{-.1em}I\hspace{-.1em}I line, while similar
shifts are obtained from the other lines. The reddening strongly
suggests that the emission lines originated from the wind-wind shock
zone, where the average velocity is $\sim600$ $\mathrm{km}$
$\mathrm{s}^{-1}$.
The red-shift motion is inconsistent
with a scenario in which the X-rays originate from the wind-wind collision
zone in the short-period binary, and
would be evidence supporting 
the widely separated O supergiant as a companion. This may make up
the collision zone be lying behind the short-period binary.
In addition to the emission lines, we also detected the RRC (radiative
recombination continuum) structure from carbon around 0.49
$\mathrm{keV}$. This implies the existence of additional cooler
plasma.
}

\authorrunning{Y.Sugawara et al. }
\titlerunning{Redshifted emission lines and RRC from the WR binary $\theta$ Muscae}

\maketitle

\section{Introduction}
Wolf-Rayet (WR) stars are the evolved descendants of massive O stars,
whose optical spectra are characterized by strong helium, carbon,
nitrogen and oxygen emission lines. These arise from hot stellar winds
with typical terminal velocities of $\approx$ 1000--3000 $\mathrm{km}$
$\mathrm{s}^{-1}$ and mass loss rates of the order of 10$^{-5}$
M$_{\sun}$ $\mathrm{yr}^{-1}$. The visible spectra of WN stars and WC
stars are dominated by nitrogen and carbon emission lines, respectively.

\citet{pollock1} used the {\it Einstein} observatory, which covered the
0.2--4.0 $\mathrm{keV}$ energy band, to discover that several WR+O
binary systems tend to be brighter in X-rays than single WR stars. This
was confirmed by a {\it ROSAT} survey \citep{pollock3}. The strong hard X-ray
emission from WR binaries can be interpreted to arise from a collision
of stellar winds from the WR star and that of the O star (e.g.,
\citealt{koyama0}). Recently, \citet{schild} discovered the RRC
(radiative recombination continuum) structure of carbon in the spectrum
of the WC+O binary $\gamma^{2}$ Velorum, which shows the existence of
another cooler component.

\begin{figure}[h]
\begin{center}
\begin{minipage}{\columnwidth}
\resizebox{0.95\columnwidth}{!}{\includegraphics[width=\columnwidth]{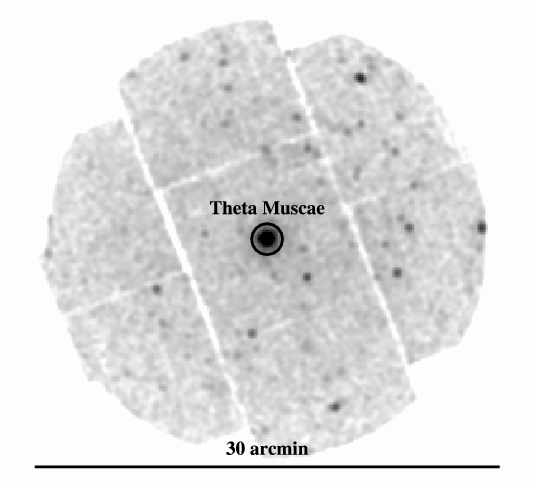}}
\caption{{\it XMM-Newton} MOS 1 image (0.3--10 $\mathrm{keV}$ band).}
\label{fig1}
\end{minipage}
\end{center}
\end{figure}

\begin{figure*}
\begin{minipage}[t]{\textwidth}
\begin{center}
\resizebox{\textwidth}{!}{\includegraphics*[angle=-90]{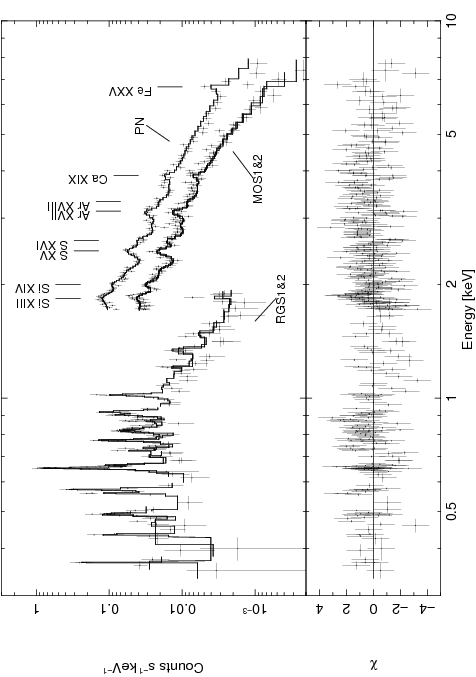}}
\end{center}
\caption{{\it XMM-Newton} RGS/MOS/PN spectra of $\theta$ Muscae. The
solid lines indicate the best-fit wabs * gsmooth * (redge $+
$ 2T vpshock) model in Table~\ref{tab1}. }
\label{fig4}
\end{minipage}
\end{figure*}

$\theta$ Muscae (WR 48, HD 113\,904) is known to be a complex
system. The history of observations and interpretations are
summarized in \citet{hill}. Based on its spectrum, this system is
known to be a binary or a line-of-sight double, and is the second brightest
WR binary in optical wavelengths \citep{moffat}. \citet{moffat} showed
that the WR star emission lines exhibited radial velocity variations
with a period of 18.341 d but the absorption lines from the late O
supergiant (O9.5/B0Iab; \citealt{hucht})
seemed to be stationary. This leads to two possible interpretations;
one is an extremely large mass ratio, and the other is the hypothesis
that the WR component shares its 18.341-d orbit with a third hidden
star, and then the O supergiant star is either in a much wider,
slower orbit with the other two or is a line-of-sight coincidence. The
linear polarization data obtained by \citet{st-louis} confirmed that
the O supergiant does not participate in the short-period
orbit. \citet{hartkopf} resolved the O supergiant to be $46\pm9$ mas away
from the short-period binary using speckle interferometry.

Non-thermal emission is detected in radio wavelengths from this system
(\citealt{leitherer}; \citealt{chapman}). \cite{dougherty} discussed, in
their study of 23 WR stars, the possibility that the non-thermal
emission arises not from a wind-wind collision zone in the short-period
binary but from that between the supergiant and the short-period
binary. The discussion is based on their obtained correlation between
orbital period and radio spectral index in WR binaries. The data point
for an orbital period of about 18 days and the spectral index of $\theta$
Mus ($\sim -$ 0.4) is offset from the correlation, while if the orbital
period is that estimated for the wide binary (at least 130 yr), the data
point lies on the correlation. The correlation would be reasonable if
we think that detection of radio non-thermal emission is difficult in
short-period systems because the wind-collision zone is deep inside the
opaque region of the stellar wind from the WR star.

In X-ray wavelengths, the {\it Einstein} observatory detected strong
X-ray emission with an observed luminosity of $2.0 \times 10^{33}$ {\rm erg
s}$^{-1}$ in the 0.2--4.0 keV band \citep{pollock1}, and the {\it ROSAT}
observatory detected emission at $1.4 \times 10^{33}$
{\rm erg s}$^{-1}$ in the 0.2--2.4 keV band \citep{pollock2}. However,
despite the past observations, the origin of the X-ray emission is not
well understood. The large effective area and the excellent spectral
resolution over a broad band of {\it XMM-Newton} allows us to measure
the detailed structure of atomic lines. We present here {\it XMM-Newton}
observations of $\theta$ Muscae.

We adopt the short orbital period of 19.1375 days is derived
most recently \citep{hill} and assume the distance to be
2.27 $\mathrm{kpc}$ \citep{hucht}.

\section{Observations and data reduction}
{\it XMM-Newton} was launched in 1999 December from French Guyana. The
observatory consists of three X-ray telescopes and five X-ray
instruments: three European Photon Imaging Cameras (EPIC: PN and two
identical MOS CCDs), which have high sensitivity up to 10
$\mathrm{keV}$, and two identical reflection grating spectrometers (RGS),
which have excellent energy resolution in the soft X-ray band (0.33--2.5
$\mathrm{keV}$).

$\theta$ Muscae was observed by {\it XMM-Newton} between 2004 July 20
11:21:43 UT and 2004 July 21 20:36:35 UT, giving 119 692 s of
observation time. For the orbital solution of the WC star, we adopted
that obtained in \citet{hill}:
\begin{equation}
\mathrm{HJD} (at\hspace{0.5em}\phi=0)= 245\,1377.51 + 19.1375 \it{E}
\end{equation}
where the eccentricity is zero and $\it{E}$ is the number of orbits
since the passage on HJD 245\,1377.51. Phase 0.0 is defined as the time
when the WR star occults its companion. Our observations cover the
orbital phase from 0.596 to 0.668, during which the O star was located
in front of the WR star. Each phase has errors $\pm$0.024 and
$\pm$0.012. The former corresponds to the uncertainty of the initial
epoch, while the latter to the offset of the orbital period.

All the data were analyzed with the 6.0.0 version of the XMM Science
Analysis System. The pipeline processing tasks EMCHAIN, EPCHAIN and
RGSPROC were executed using the available calibration files, and data were
filtered via EVSELECT to select good event patterns. We use all valid
event patterns (PATTERN 0--12) for the MOS cameras but use valid PN
events with single and double events (PATTERN 0--4).
In the energy band below 1.8 $\mathrm{keV}$, the emission-line peaks
of the EPIC spectrum are not consistent with those of the high
resolution RGS spectrum. We adopt the RGS spectrum
and ignored the EPIC spectrum in this energy band
to avoid the statistical dominance by EPIC data.

\section{Results}

Figure~\ref{fig1} shows the X-ray image taken with MOS 1. $\theta$
Muscae is the dominant X-ray source in the MOS field of view. We derived
light curves in the soft band (0.3--2 $\mathrm{keV}$) and hard band
(2--10 $\mathrm{keV}$). The average count rates are 0.13 and $3.1 \times
10^{-2}$ {\rm counts s}$^{-1}$ respectively. We detected no significant
variability.

\begin{figure*}[t]
\begin{minipage}{\textwidth}
\begin{center}
\resizebox{0.86\textwidth}{!}{\includegraphics*[angle=-90]{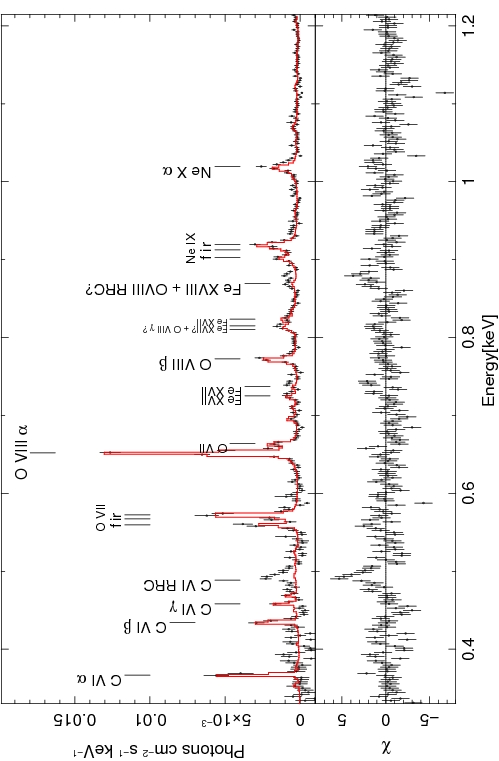}}
\resizebox{0.86\textwidth}{!}{\includegraphics*[angle=-90]{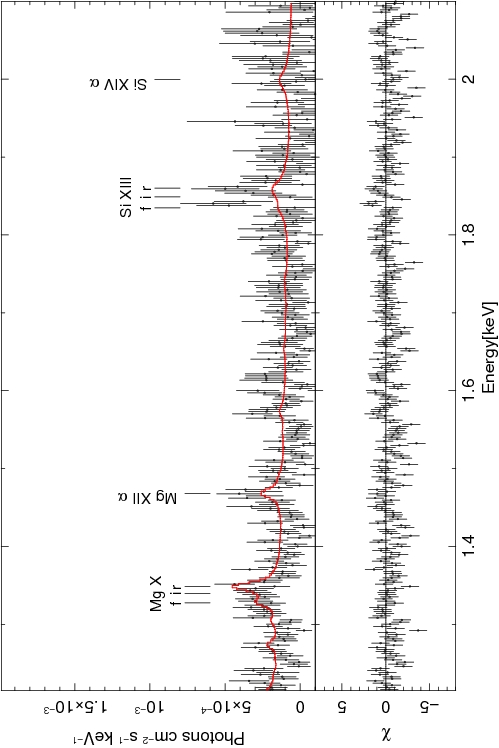}}
\caption{Unfolded {\it XMM-Newton} RGS spectrum of $\theta$
Muscae made with SAS task rgsfluxer. The
solid lines indicate wabs * gsmooth * 2T vpshock model. The RRC model (redge
model) is not included in the model (see text).}
\label{fig3}
\end{center}
\end{minipage}
\end{figure*}

\begin{table*}
\begin{minipage}[t]{\textwidth}
\caption{Results of spectral fitting with the
multi-temperature models. Errors and upper limits are at 90\% confidence
level.}
\label{tab1}
\renewcommand{\footnoterule}{}
\begin{center}
\begin{tabular}{l l c c}
Model & & wabs * gsmooth * (redge $+$ cevmkl) & wabs * gsmooth * (redge $+
$ 2T vpshock)\vspace{1mm}\\ \hline\hline
\vspace{-3mm}\\
& Parameter & Best-fit values & Best-fit values\vspace{1mm}\\
\hline
\vspace{-3mm}\\
\multicolumn{4}{l}{Line of sight absorption (wabs)} \\
& $N_{\mathrm{H}}$ (10$^{21}$ $\mathrm{cm}^{-2}$)
&2.56$^{+0.07}_{-0.06}$&2.4$^{+0.2}_{-0.3}$\vspace{1mm}\\
\multicolumn{4}{l}{Doppler broadening (gsmooth)} \\
& $\sigma_{6}$ \footnote{$\sigma_{6}$ is the standard deviation at 6
$\mathrm{keV}$.The broadening corresponds to a velocity of 1410 $\mathrm{km}$ $\mathrm{s}^{-1}$ in FWHM.} ($\mathrm{eV}$) &\multicolumn{2}{c}{11.95 (fixed)}
\vspace{1mm}\\
& Index \footnote{This is the power of energy for sigma variation.}& \multicolumn{2}{c}{1 (fixed)}\vspace{1mm}\\
\multicolumn{4}{l}{Doppler shift (redshift)} \\
& Redshift ($\times$ 10$^{-3}$) &2.18$^{+0.08}_{-0.26}$&2.17$^{+0.16}_{-0.06}$\vspace{1mm}\\
\multicolumn{4}{l}{RRC (redge)} \\
& Threshold energy ($\mathrm{eV}$)& 489.2$^{+2.8}_{-0.9}$ &
489.2$^{+1.0}_{-0.9}$ \vspace{1mm}\\
& $kT_{\rm RRC}$ ($\mathrm{eV}$)& 4$\pm 1$ &5$^{+2}_{-1}$
\vspace{1mm}\\
& norm \footnote{Normalization constant of the redge model defined as
total photons ${\rm cm}^{-2}$ s$^{-1}$ in the line.}($\times$
10$^{-4}$)& 1.5$\pm 0.3$ & 1.1$\pm 0.2$ \vspace{1mm}\\
\multicolumn{2}{l}{Thin thermal plasma model}
& (cevmkl) & (2T vpshock)\\
& $\alpha$ & $-0.25\pm 0.02$&--
\vspace{1mm}\\
& $kT_{max}$ ($\mathrm{keV}$) &
9.9$^{+0.7}_{-0.5}$&--\vspace{1mm}\\
& $kT_{1}$ ($\mathrm{keV}$)&--&0.58$^{+0.03}_{-0.12}$
\vspace{1mm}\\
& $kT_{2}$ ($\mathrm{keV}$)&--&3.6$\pm 0.1$\vspace{1mm}\\
& Abundance \footnote{The solar abundances are from
\citet{anders}. For the elements with no emission lines (hydrogen,
helium, sodium and aluminum), the solar abundances are kept
fixed. The abundance of magnesium and nickel are linked to that of
silicon and iron respectively.} &&
\vspace{1mm}\\
& \hspace{2mm} C
&14.8$^{+3.4}_{-2.7}$&3.5$^{+1.8}_{-0.6}$\vspace{1mm}\\
& \hspace{2mm} N &$< 10^{-8}$ &$<$ 0.1\vspace{1mm}\\
& \hspace{2mm} O &1.9$\pm 0.1$&0.33$\pm 0.09$
\vspace{1mm}\\
& \hspace{2mm} Ne &1.8$\pm 0.3$&0.47$^{+0.10}_{-0.06}$
\vspace{1mm}\\
& \hspace{2mm} Si &1.0$\pm 0.1$&0.31$^{+0.05}_{-0.02}$
\vspace{1mm}\\
& \hspace{2mm} S &2.2$\pm 0.2$&0.64$^{+0.10}_{-0.07}$
\vspace{1mm}\\
& \hspace{2mm} Ar &4.5$^{+0.8}_{-0.7}$&1.1$\pm 0.2$
\vspace{1mm}\\
& \hspace{2mm} Ca &3.7$\pm 1.1$&1.0$^{+0.4}_{-0.3}$
\vspace{1mm}\\
& \hspace{2mm} Fe &0.21$\pm
0.03$&0.15$^{+0.02}_{-0.03}$\vspace{1mm}\\
& $\tau_{u}$($\times$ 10$^{11}$ s cm$^{-3}$)&--&4.9 \vspace{1mm}
\\

& norm$_{1}$ \footnote{\label{fn1} Normalization constant
defined as
$\mathrm{EM} \times 10^{-14}(4 \pi \mathrm{D}^{2})^{-1}$,
where
$\mathrm{EM}$ is the emission measure in $\mathrm{cm}^{-3}$ and
$\mathrm{D}$ is the distance in $\mathrm{cm}$. In the cevmkl model,
emission measure follows a power law in temperature, i.e. the emission
measure from electron temperature {\it kT} is proportional to $\left({\it kT}/{\it kT}_{max}\right)^{\alpha}$.}
($\times$10$^{-4}$)
&4.4$\pm 0.1$&5.8$^{+0.7}_{-1.6}$
\vspace{1mm}\\
& norm$_{2}$ $^{\it e}$ ($\times$ 10$^{-4}$)&--&8.3$^{+0.5}_{-0.3}$\vspace{1mm}
\\
& $L_{\mathrm{X}}$ \footnote{The absorption-corrected luminosity
(0.3--8.0 {\rm keV}) was calculated assuming a distance of 2.27
{\rm
kpc}.} ($\mathrm{erg}$ $\mathrm{s}^{-1}$ )&3.6 $\times$
10$^{33}$&3.1 $\times$ 10$^{33}$\vspace{1mm}\\
\hline
\vspace{-3mm}\\
& $\chi^{2}$/d.o.f &762/429&605/426

\end{tabular} 
\end{center}
\end{minipage}
\end{table*}

\begin{table*}
\begin{minipage}[t]{\textwidth}
\caption{Doppler shift of emission lines and RRC. Errors are at 90\% confidence level.}
\label{tab0}
\renewcommand{\footnoterule}{}
\begin{center}

\begin{tabular}{c c c c c c c c c c c}
\multicolumn{2}{l}{} \\ \hline\hline
\vspace{-3mm}\\
Ion \footnote{Doppler shifts of the emission lines were measured with
a single Gaussian fitting. That of the RRC was picked from the best-fit
parameter of the 2T vpshock model in Table~\ref{tab1}.} & C V\hspace{-.1em}I & C V\hspace{-.1em}I & O
V\hspace{-.1em}I\hspace{-.1em}I & O
V\hspace{-.1em}I\hspace{-.1em}I\hspace{-.1em}I & Ne IX & Ne X & C RRC
\vspace{1mm} \\ \hline
$E_{0} ({\rm keV})$\footnote{Theoretical energy of the line center.} & 0.3675 & 0.4356 & 0.5739 & 0.6536 & 0.9050 &
1.0218 & 0.4900 \vspace{1mm}\\
$E_{\rm{obs}} ({\rm keV})$\footnote{The center energy of the observed line.} & 0.3668 & 0.4344 & 0.5725 & 0.6522 & 0.9022 &
1.0196 & 0.4892 \vspace{1mm}\\
Redshift \footnote{The redshift defined as $(E_{0}-E_{\rm{obs}})/E_{\rm{obs}}$.} ($\times$ 10$^{-3}$) & 1.9$^{+0.9}_{-1.9}$ &
2.8$\pm 0.6$ & 2.5$\pm 0.5$ & 2.1$\pm 0.3$ &
3.1$^{+1.1}_{-1.5}$ & 2.2$^{+0.8}_{-0.9}$ & 1.6$^{+1.8}_{-2.0}$ \vspace{1mm}\\
$v$ ($\mathrm{km}$ $\mathrm{s}^{-1}$) &560$^{+270}_{-560}$ & 830$^{+170}_{-180}$ & 750$^{+140}_{-150}$ & 630$^{+80}_{-100}$ & 930$^{+320}_{-440}$ &
650$^{+240}_{-260}$ & 490$^{+550}_{-611}$ \vspace{1mm}\\
\hline
\end{tabular}
\end{center}
\end{minipage}
\end{table*}

We obtained X-ray spectra from the RGS and EPIC (PN, MOS 1 and 2) data. The
broad band spectra are shown in Figure~\ref{fig4} and the enlarged
spectrum for the RGS data is shown on a linear scale in
Figure~\ref{fig3}. Emission lines from helium-like and hydrogen-like
ions of various elements (e.g. carbon, oxygen and neon) were
detected. Since a single temperature plasma cannot reproduce these
emission lines simultaneously, we executed simultaneous fitting of the
three datasets (RGS, MOS and PN) with a multi-temperature plasma model
with an absorption model (wabs) using XSPEC ver. 12. The
multi-temperature model we adopted first is the simple collisional
ionization equilibrium (CIE) model (cevmkl), where the emission measures
follow a power-law in temperature, i.e, the emission measure from the electron
temperature {\it kT} is proportional to $\left({\it kT}/{\it
kT}_{max}\right)^{\alpha}$ with an index $\alpha$ \citep{singh}. The
secondary adopted model is the two temperature plane-parallel shock
model (2T vpshock), which does account for non-equilibrium ionization
(NEI).

In both fittings, line broadening and shifts were seen. Then we fitted
the high-intensity O V\hspace{-.1em}I\hspace{-.1em}I\hspace{-.1em}I
$\alpha$ line around 0.65 {\rm keV} with a single Gaussian model to
investigate the width. The obtained broadening is $1.3 ^{+0.3}_{-0.2}$ {\rm
eV} in 1 $\sigma$, which corresponds to a velocity of 1190--1720
$\mathrm{km}$ $\mathrm{s}^{-1}$ in FWHM. With this value, we adopted a Gaussian
smoothing (gsmooth) of the multi-temperature models, fixing the ratio
$\sigma / E \sim 1.3$ eV $/$ 0.65 keV and allowing the redshift to be a free
parameter.

The above fittings reasonably reproduced the overall spectrum. The
solid line in Figure~\ref{fig3} indicates the model for NEI. However,
residuals around 0.49 {\rm keV} were left (see Figure~\ref{fig3}) in the
RGS band. The excess is identified with the RRC 
structure of C VI. We then added a recombination edge (redge) model
to the above models, and re-fitted the datasets. The results are shown
in Table~\ref{tab1}. The better fit is obtained with the NEI model. The
solid lines in Figure~\ref{fig4} show the best-fit NEI model.

The best-fit NEI model, and also the CIE model, showed that carbon is
overabundant by at least an order of magnitude compared to nitrogen in
$\theta$ Muscae (see Table~\ref{tab1}), indicating that the plasma
originates from Wolf-Rayet (WC) stellar winds. The best fit model has
an absorption-corrected X-ray luminosity ($L_{\mathrm{X}}$) of $3.1 \times
10^{33}$ {\rm erg s}$^{-1}$ (0.3--8.0 {\rm keV} band) and log
$L_{\mathrm{X}}/L_{\rm{bol}}$ of $-5.56$ (log $L_{\rm{bol}}/L_{\sun} =
5.47$; \citealt{nugis}), which is slightly larger than the known
correlation $L_{\mathrm{X}} = 10^{-7} L_{\rm{bol}}$ for O+O and WN+O
binaries (\citealt{berghoefer}; \citealt{oskinova}). The
absorption of $2.4 \times 10^{21}$ {\rm cm}$^{-2}$ is consistent, within
a factor of 1.5, to the interstellar absorption $1.6\times 10^{21}$
$\mathrm{cm}^{-2}$ derived from $A_{V} = 0.93$ {\rm mag} and the
correlation in \cite{predehl}. The red-shift is 2.2
$\times$ $10^{-3}$, which corresponds to 650 km
s$^{-1}$. Independently, we further fitted each emission line with a
single Gaussian model, and obtained a similar red-shift $\sim$ 2 $\times$
$10^{-3}$ (see Table~\ref{tab0}).

The RRC structure of carbon we detected is the second example among WR
massive binaries, with the first case being the spectrum of
$\gamma^{2}$ Velorum \citep{schild}. The fitting of the RRC structure in
$\theta$ Muscae needed a plasma temperature of 4.7 (3.3--6.2) {\rm
eV}, which is similar to but about 1.3 times higher than that of
$\gamma^{2}$ Velorum. The cool plasma component which makes up the RRC
structures may be common in WR binaries. Interestingly, the best-fit
value of the RRC indicated a red-shift of 2.5$\times10^{-3}$, which
is similar to that obtained in lines, although the error is large.

We further analyzed the emission lines around 0.56 {\rm keV} and 0.9
{\rm keV}, which are the helium-like triplets of oxygen and neon,
respectively. The intensity ratio of forbidden lines (HE6) to resonance
lines (HE4) indicates that the lines originate from a thin-thermal
plasma whose density is less than $10^{11} \mathrm{cm}^{-3}$
\citep{porquet}, which favors the NEI model.

\section{Discussion}

The X-ray spectra from the massive binaries WR140 \citep{pollock4},
$\gamma^{2}$ Velorum \citep{schild}, WR25 \citep{raassen}, V444 Cygnus
\citep{maeda} are all strong in hard X-rays with high temperature
components of several keV. The hot component of $\theta$ Muscae also strongly
suggests that X-rays originate from wind-wind collisions. The
absorption-corrected X-ray
luminosity of $\theta$ Muscae is 3.1 $\times$ 10$^{33}$ {\rm erg
s}$^{-1}$, which is similar to those of the other massive binaries
(WR140, 2.0 $\times$ 10$^{34}$ {\rm erg s}$^{-1}$; $\gamma^{2}$ Velorum,
8.4 $\times$ 10$^{32}$ {\rm erg s}$^{-1}$; WR25, 1.3 $\times$ 10$^{34}$
{\rm erg s}$^{-1}$; V444 Cygnus, 1.4 $\times$ 10$^{33}$ {\rm erg
s}$^{-1}$ ).

High dispersion spectra of the massive binaries have been obtained for
WR140 \citep{pollock4}, $\gamma^{2}$ Velorum \citep{henley}, and
$\theta$ Muscae (this study). Significant Doppler shifts are only reported by
\citet{pollock4} for WR140 at the pre-periastron phase when one side of
the bow-shock zone approaches toward the line of sight. They
detected a significant blue shift of 600 $\mathrm{km}$ $\mathrm{s}^{-1}$. No
detection of the Doppler shift at post-periastron was interpreted to
result from the bow-shock zone being aligned perpendicular to the line of sight.
No detection from $\gamma^{2}$ Velorum is also attributed to a
similar situation or to a wide open angle of the bow-shock zone.

\begin{figure}[t]
\begin{center}
\begin{minipage}{\columnwidth}
\resizebox{0.9\columnwidth}{!}{\includegraphics[width=\columnwidth]{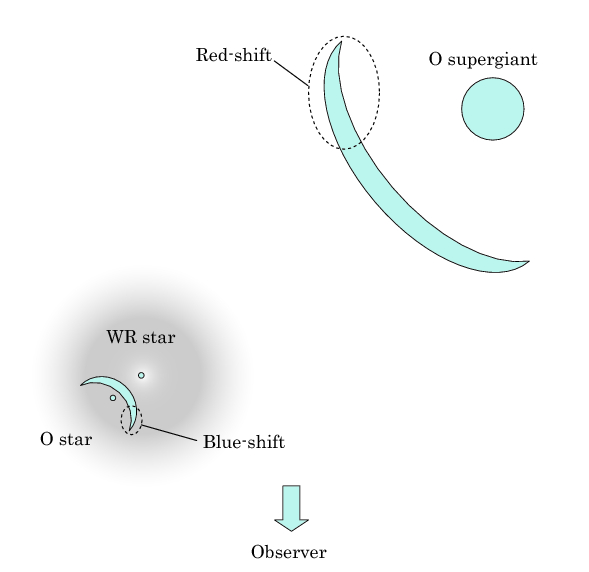}}
\caption{Schematic image of $\theta$ Muscae.}
\label{test}
\end{minipage}
\end{center}
\end{figure}

Our detection of the Doppler-shift in the emission lines is the second
example among the massive binaries. The red-shift in the emission lines
strongly suggests that one side of the bow-shock zone is receding along the
line of sight, indicating that the primary WR star is in front. However, this
geometry is inconsistent with the orbital solution for the short-period
binary in \citet{hill} in which the O star is likely in front.
Instead, if the O supergiant separated by 46~mas from the
short-period binary is a companion of this system and
located behind the primary WR star,
as shown in Figure~\ref{test}, the wind-wind collision zone could be receding.

This idea that the wind-wind collision zone is between the short-period
binary and the O supergiant is the same as that of \cite{dougherty}.
By assuming that the distance to $\theta$ Muscae is 2.27 kpc, the projection
of 46~mas corresponds to $\sim$100~AU. It is well known that X-rays
from massive binaries suffer absorption from the WR wind
if the WR star is in front (e.g., \citealt{maeda}). An X-ray emitting
region widely extended over 100~AU should avoid the absorption in the
$\theta$ Muscae system. This interpretation strongly suggests that
the astrometric O supergiant is a companion to the system. The wind from the
spectroscopic O star companion should be much weaker than that from the
optically identified O supergiant.

Another interesting possibility was obtained from the RRC structure. 
Table~\ref{tab0} shows that the RRC may be red-shifted with a similar
velocity to the emission lines. If this is true, the RRC may originate
from ions that escape from the bow-shock layer through diffusion or
convection processes, and exchange electrons in the following
less-ionized winds. Follow-up observations with much deeper exposure
are necessary to test this idea since we need to restrict the Doppler
shift of the RRC structure more tightly.

\section{Summary}

The {\it XMM-Newton} spectra of $\theta$ Muscae show He- and H-like
emission lines from various elements as well as the RRCs from carbon.
The results of our study are as follows:
\begin{itemize}
\item [A.] The He-like and H-like emission can be reproduced by a
      multi-temperature plasma model. A better fit was obtained for
      the NEI case (vpshock) rather than the collisional equilibrium
      (cevmkl) model. The high temperature component of the NEI model is
      as high as $kT\sim$3 {\rm keV}. The high temperature component
      indicates the existence of plasma heating by a wind-wind collision
      shock.\\

\item [B.] We detected RRC structure which implies the existence of a
      cooler component of around $kT\sim5~{\rm eV}$.\\

\item [C.] The abundance of carbon is at least one order of magnitude
      higher than that of nitrogen. The over-abundance indicates that
      the Wolf-Rayet (WC) stellar winds dominate the X-ray emitting gas.\\

\item [D.] The emission lines from carbon, oxygen, neon and silicon and
      possibly the RRC from carbon show red-shifts of $\sim600$
      $\mathrm{km}$ $\mathrm{s}^{-1}$ with broadenings of $\sim1400$
      $\mathrm{km}$ $\mathrm{s}^{-1}$ in FWHM. The red-shift would be
      evidence supporting the widely separated O supergiant as a
      companion with which the collision zone could be formed, lying
      behind the short-period binary. The wind from the O star in the
      short-period binary should be much weaker than that from the O
      supergiant in the wide binary.\\

\end{itemize}

\begin{acknowledgements}
We thank A.Yamamoto for analysis and technical support. We thank
C.Baluta for English correction. T.Tsujimoto gave us valuable comments
on the chemical evolution of massive stars. Y.\,T. and
Y.\,M. acknowledge support from the Grants-in-Aid for Scientific
Research (numbers 20540237 and 20041010) by the Ministry of Education, Culture, Sports, Science and Technology.

\end{acknowledgements}

\end{document}